\documentclass[]{iopart}
\usepackage{amssymb}
\usepackage{graphics}
\usepackage{cite}

\newcommand{\mrm}[1]{\mathrm{#1}}

\begin{document}


\title[Target mass number dependence of subthreshold antiproton production]
	{Target mass number dependence of subthreshold antiproton
        production in proton-, deuteron- and alpha-particle-induced reactions}

\author{H M\"uller$^1$ and V I Komarov$^2$}
\address{$^1$ Institut f\"ur Kern- und Hadronenphysik,
        Forschungszentrum Rossendorf, Postfach 510119, D-01314 Dresden,
        Germany}
\address{$^2$ Joint Institute for Nuclear Research, LNP, 141980 Dubna,
	Russia}
\ead{H.Mueller@fz-rossendorf.de}

\begin{abstract} Data from KEK on subthreshold $\bar{\mrm{p}}$ as well
  as on $\pi^\pm$ and $\mrm{K}^\pm$~production in proton-, deuteron- and
  $\alpha$-induced reactions at energies   between 2.0 and 12.0~A GeV
  for C, Cu and  Pb targets are described  within a unified  approach.
  We use  a model which considers a  nuclear reaction as an incoherent
  sum over collisions of   varying numbers  of projectile  and  target
  nucleons.  It samples complete   events   and thus allows  for   the
  simultaneous consideration of   all final  particles including   the
  decay products of the nuclear residues.  The enormous enhancement of
  the $\bar{\mrm{p}}$ cross section,  as well as the moderate increase
  of  meson production  in  deuteron and $\alpha$  induced compared to
  proton-induced reactions, is  well reproduced for all target nuclei.
  In  our approach,  the   observed enhancement  near the   production
  threshold is  mainly due to  the contributions from the interactions
  of few-nucleon clusters  by simultaneously considering fragmentation
  processes of  the  nuclear residues.   The  ability of  the model  to
  reproduce the target mass dependence may be  considered as a further
  proof of the validity of the cluster concept.
\end{abstract}

\submitto{\JPG}

%

\section{Introduction\label{intro}}

In   two   recent  papers   \cite{komarov04,muellerh04} we   discussed
data~\cite{sugaya98}      on  $\bar{\mrm{p}}$,    $\pi^\pm$        and
$\mrm{K}^\pm$~production in the  reactions  pC,  dC and $\alpha$C   at
energies  between 3.5 and 12.0~A GeV  within  the Rossendorf collision
(ROC)   model.  Special emphasis   was on  subthreshold production  of
antiprotons the   enormous enhancement of  which  in  dC and $\alpha$C
compared  to pC reactions could be  well reproduced. Here, we consider
the  target mass number dependence of  this effect  by confronting the
KEK data~\cite{sugaya98} on  hadron production in pCu, dCu, $\alpha$Cu
and dPb reactions with ROC-model calculations and make predictions for
the reactions pPb and $\alpha$Pb not measured so far.

\section{The model\label{model}}
The    ROC    model  considers    a   nuclear   reaction   within  the
spectator-participant picture as an incoherent  sum over collisions of
varying  numbers  of projectile   and target  nucleons. Partial  cross
sections for  these interactions are    derived from a   probabilistic
interpretation of the Glauber theory~\cite{glauber70} in close analogy
to the cooperative model
\cite{knoll79,knoll80,bohrmann81,shyam84,shyam86,knoll88,ghosh90,ghosh92}.
They depend strongly on the mass numbers  via the wave functions of the
nuclei. For light nuclei they drop more rapidly with increasing number
of participants than in case of heavier nuclei.

Particle  production  is proposed  to  proceed via intermediate states
called fireballs (FBs).  This assumption  yields a natural explanation
of   particle  correlations  observed  at    higher energies (see  the
discussion in  \cite{muellerh01}).   Two parameters,  temperature  and
radius, define number and relative kinetic energy of the final hadrons
the FBs decay into.  Resonances among the final hadrons decay later on
into stable particles observed in the  experiment. The nuclear residue
is assumed to become excited during the reaction due to the distortion
of the nuclear  structure by the separation  of the  participants from
the spectators and due to the passage of the reaction products through
the spectator system.  In this way, final-state interactions are taken
into     account  without   making    special assumptions   concerning
re-absorption,  re-scattering, self-energies, potentials  etc for each
particle    type    separately. Obviously,   a    comparison  of model
calculations  with data    for heavier  nuclei  should   be  much more
sensitive to the treatment of such secondary effects.

As thoroughly discussed in \cite{komarov04},  the consideration of the
fragmentation  of the  nuclear  residues is   decisive for reproducing
subthreshold  cross sections.  The  model  describes fragmentation  in
analogy to  the decay of the  excited FBs also by  a temperature and a
radius  parameter.   Obviously, the  main   ingredients of  the  model
(partial cross  sections, FB excitation,  final-state interactions and
fragmentation)  may influence  the   total mass number  dependence  of
hadron production   cross  sections  quite  differently.    Thus,  the
consideration of the mass  number dependence is  an important check to
verify the model assumptions.

The   mathematical formulation of  the  above   picture  is given   in
\cite{muellerh04}.  In the following  we employ the same parameter set
as used  in \cite{komarov04,muellerh04} for the C  target also for the
heavier  target nuclei except the  maximum  temperature of the nuclear
residues (see section \ref{comp}).

\section{Comparison with experimental data \label{comp}}
\begin{figure}
\begin{center}
   \resizebox{0.98\textwidth}{!}{
   \includegraphics*{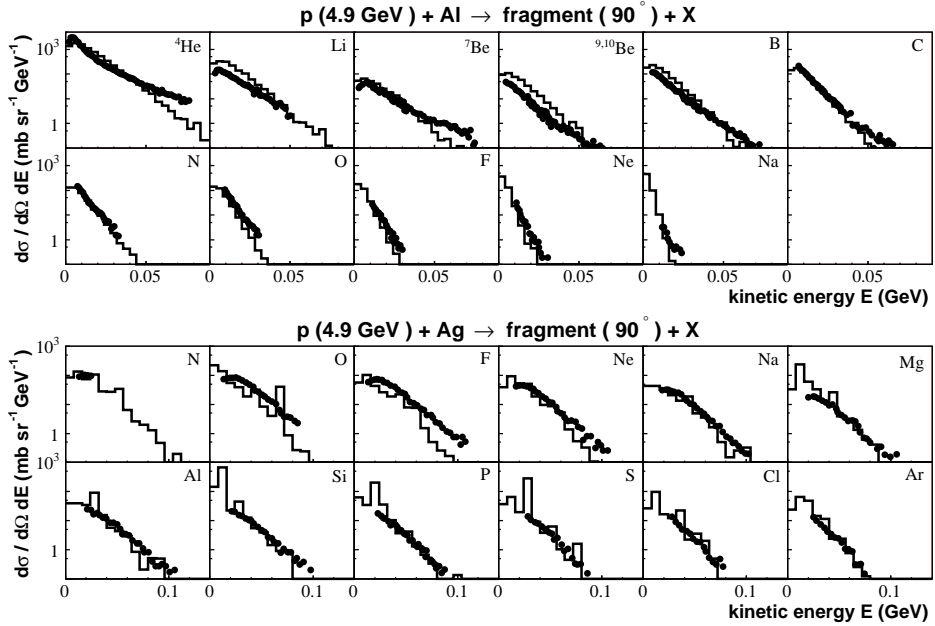}}
\end{center}

   \caption{\label{p4.9A}    Energy  spectra \cite{westfall78}  of the
   nuclear   fragments (symbols) indicated    in the figure from  p+Al
   (upper panel) and p+Ag  (lower panel) reactions at 4.9~GeV compared
   with     ROC-model   calculations  (histograms)               using
   $\Theta_R^{max}$=10~MeV for  Al   and  $\Theta_R^{max}$=6.8~MeV for
   Ag.}
\end{figure}
As  discussed  in \cite{muellerh92,komarov04}, one  of the outstanding
features   of the ROC   model is  the   unified treatment  of particle
production and fragmentation  processes.   The interplay of  these two
energy consuming processes is  of  special importance at  subthreshold
energies  where  the   available  energy   is decisive   for  particle
production  cross sections.  Therefore, we  start our consideration of
mass number dependence of (subthreshold) particle production by fixing
the parameters of  the  nuclear residues from   independent data
\cite{katcoff68,english74,kaufman76,kaufman80,westfall78}          for
multifragmentation  in  proton-nucleus interactions.   For simplicity,
the    radius parameter  $R_R=1.7  \,  \mrm{fm}$   and  the  value  of
$\bar{a}=0.5$ (equation~(17)  in \cite{muellerh04}) defining       the
dependence of the temperature  parameter $\Theta_R$ from the number of
participants $a$
\[
   \Theta_R(a,A) = \Theta_R^{max}\, [1-\exp{(-a/\bar{a}A^{1/3})]}
\]
are taken as constant.  A  reasonable overall reproduction of the data
could  be achieved  with $\Theta^{max}_R$ taken   as a function of the
target mass  number,  which strongly decreases  at  light  and becomes
constant  at heavy target masses.   The fragment spectra result from a
rather  complicated  superposition of  contributions   with  different
values of $\Theta_R(a,A)$, which are for small numbers of participants
much  smaller  than    $\Theta^{max}_R$.    The  higher  values     of
$\Theta^{max}_R$ at low residue  mass numbers do not necessarily imply
a higher 'effective'  $\Theta_R$ contributing to the  spectra, because
for light  nuclei the  partial  cross sections  drop more rapidly with
increasing number   of participants than  in  case of  heavier nuclei.
Anyway,  decisive for the  intended discussion of hadron production is
the reproduction of  existing data to  get a realistic estimate of the
energy consumed by fragmentation.

As  example,  figure~\ref{p4.9A} demonstrates that   the basic
features  of    fragment  spectra   are  reasonably  reproduced.    In
\cite{muellerh04}, we used $\Theta^{max}_R$=12~MeV for light nuclei up
to   C,    here $\Theta^{max}_R$=10~MeV    is     taken  for   Al  and
$\Theta^{max}_R$=6.8~MeV for  Ag (see  figure \ref{p4.9A}).  For  mass
numbers above 100  the temperature is assumed  to  be constant  and we
take $\Theta^{max}_R$=6.8~MeV also for Pb.  Since in the region around
Cu there are no data available we (nonlinearly \footnote{All values of
$\Theta^{max}_R$     used  are      approximately  described        by
$\Theta^{max}_R/\mrm{MeV}=6.8[1+1.2\, \exp(-0.036\,A)]$  with  A being
the    target    mass      number.})      interpolate     and      get
$\Theta^{max}_R$=7.5~MeV  for Cu.  It should  be stressed that for the
following considerations  all other model  parameters  are the same as
used in
\cite{komarov04,muellerh04} for describing the experimental results of
\cite{sugaya98} for interactions with carbon.

\begin{figure}
\begin{center}
   \resizebox{0.98\textwidth}{!}{
   \includegraphics*{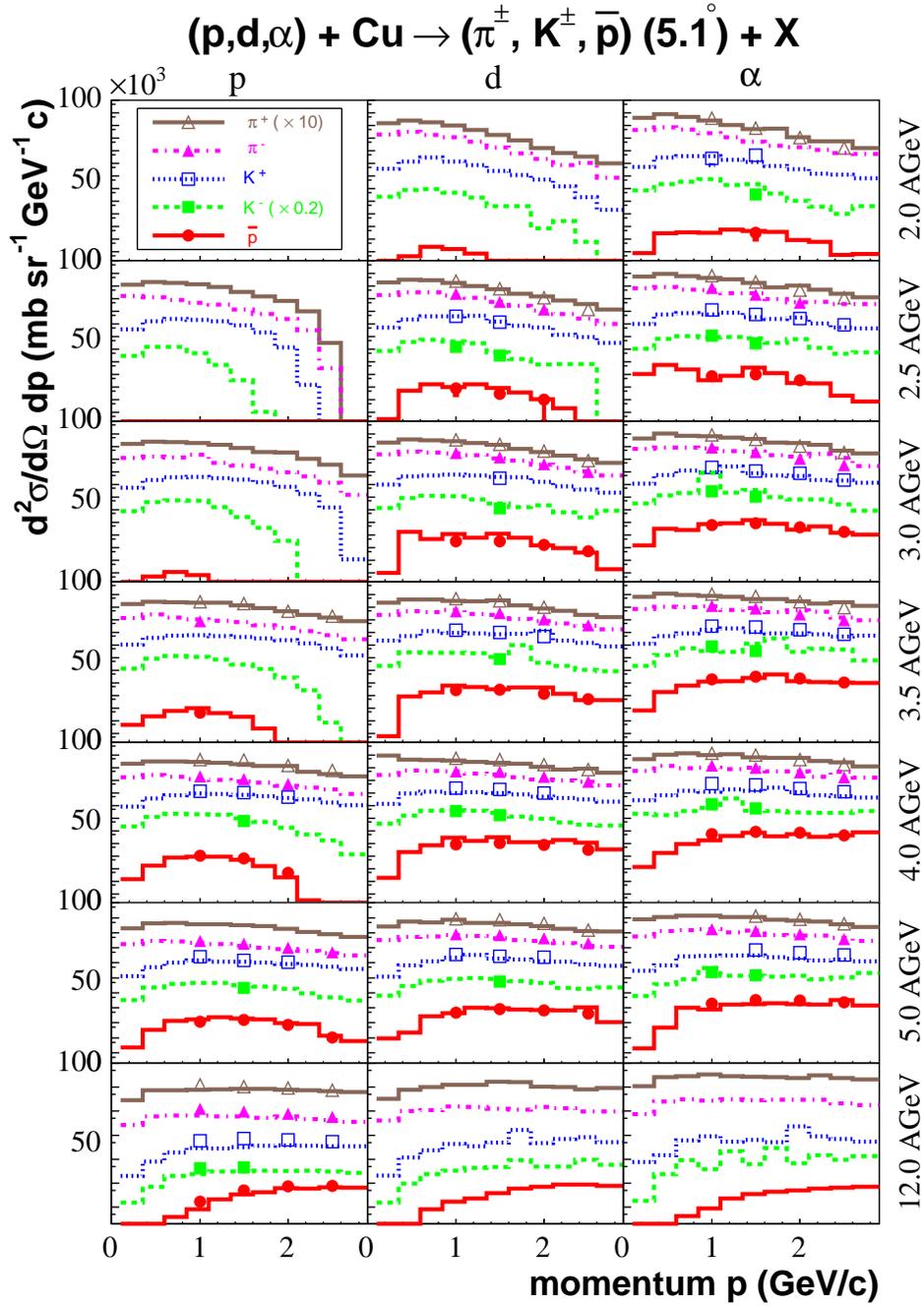}}
\end{center}
   \caption{\label{fCuAll}   Momentum   spectra        of   $\pi^\pm$,
   $\mrm{K}^\pm$ and  $\bar{\mrm{p}}$ \cite{sugaya98}  (symbols)  from
   pCu,   dCu   and  $\alpha$Cu   reactions compared    with ROC-model
   calculations (histograms).  Experimental and calculated results for
   $\pi^+$ and  $\mrm{K}^-$  mesons are   multiplied  by the   factors
   indicated in the legend.  The scale extends from $2 \times 10^{-8}$
   to $10^5 \, \mrm{mb \, sr^{-1} \,  GeV^{-1}\,c}$ for energies up to
   4.0~A GeV, while  the lower limit is  $2 \times 10^{-6}$ for  5.0~A
   GeV and $2 \times 10^{-2} \, \mrm{mb \, sr^{-1}\, GeV^{-1}\,c}$ for
   12.0~A GeV.}
\end{figure}

In   figure~\ref{fCuAll},  the     momentum   spectra  of   $\pi^\pm$,
$\mrm{K}^\pm$  and  $\bar{\mrm{p}}$   from  pCu,  dCu  and  $\alpha$Cu
reactions   obtained   at  KEK~\cite{sugaya98}  are   displayed.  This
experiment was  motivated by measurements of  large  cross sections of
subthreshold $\bar{\mrm{p}}$~production in nucleus-nucleus   reactions
at  LBL-BEVALAC~\cite{caroll89,shor89} and at  GSI~\cite{schroeter93}.
The  authors~\cite{sugaya98} claimed that the  use  of light-ion beams
for investigating subthreshold  $\bar{\mrm{p}}$  production should  be
useful  to   verify  theoretical   models.    Secondary  effects  like
re-scattering,  re-absorption, self-energies,  potentials  etc, which
hide and influence features  of the primary production process, should
be smaller than  in heavy-ion reactions.  Thus,  it is the outstanding
feature of   the data  that in  d-   and $\alpha$-induced reactions an
enormous enhancement of  $\bar{\mrm{p}}$~production by nearly  two and
three orders of magnitude compared to  p-induced interactions could be
observed.   This   enhancement makes   light-ion-induced  reactions  a
candidate   for the  key  to a   deeper understanding  of subthreshold
$\bar{\mrm{p}}$~production in nucleus-nucleus reactions.
\begin{figure}
\begin{center}
   \resizebox{0.98\textwidth}{!}{
   \includegraphics*{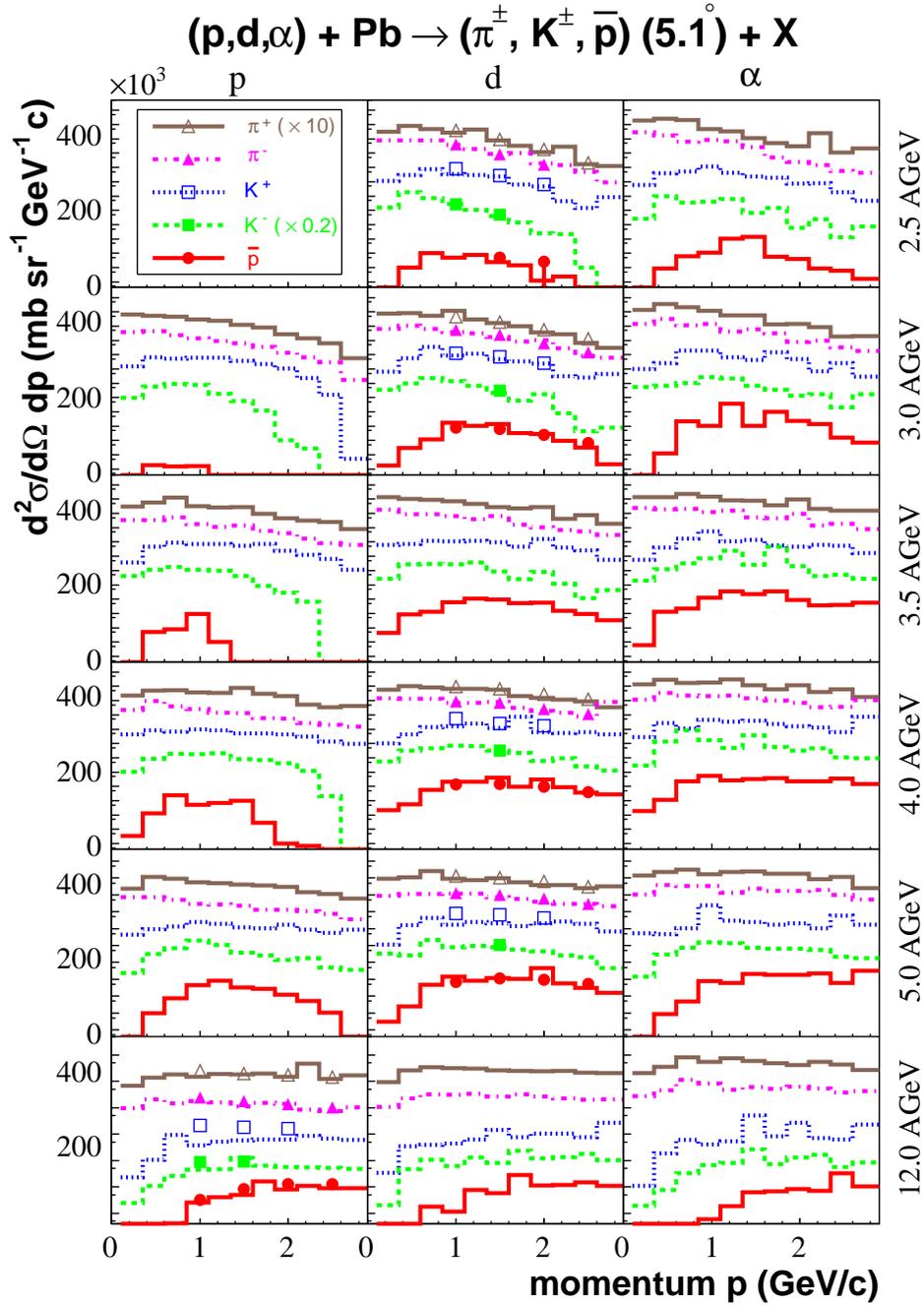}}
\end{center}
   \caption{\label{fPbAll}  The same  as   figure \ref{fCuAll} for  Pb
   target. The scale extends from $4 \times 10^{-8}$ to $5 \times 10^5
   \, \mrm{mb  \, sr^{-1}  \, GeV^{-1}\,c}$ for  energies  up to 4.0~A
   GeV, while the lower limit is $4 \times  10^{-6}$ for 5.0~A GeV and
   $4 \times 10^{-2} \,  \mrm{mb \, sr^{-1}\, GeV^{-1}\,c}$ for 12.0~A
   GeV.}
\end{figure}
\begin{figure}
\begin{center}
   \resizebox{0.98\textwidth}{!}{
   \includegraphics*{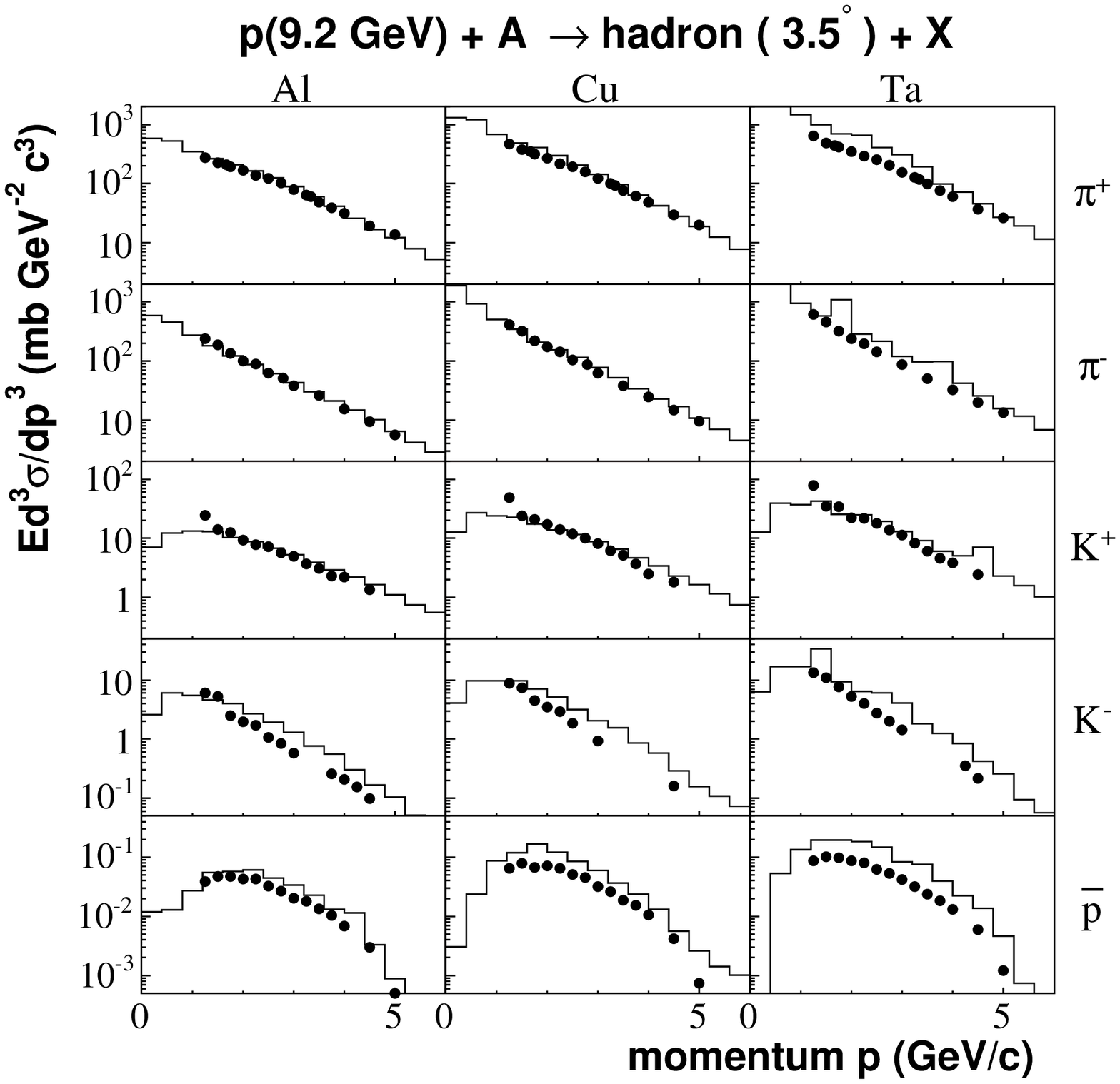}}
\end{center}
   \caption{\label{ITEP}  Momentum spectra  of   $\pi^\pm$, $K^\pm$ and
   $\bar{p}$   \cite{vorontsov88} (symbols)    compared to   ROC-model
   calculations (histograms).}
\end{figure}

The KEK group~\cite{sugaya98} interprets their $\bar{\mrm{p}}$~data by
using the `first-chance  NN collision model' from \cite{shor90}  where
the internal nucleon momenta   were  extracted from  backward   proton
production \cite{geaga80}   as   a   superposition of  two    Gaussian
distributions. Momentum spectra and the incidence-energy dependence of
p-induced reactions could  be successfully reproduced  by adapting one
normalization parameter.  However,  the  application of this  model to
d-induced reactions severely  underestimates the $\bar{\mrm{p}}$ cross
sections    at  subthreshold    energies.     Thus,    according    to
\cite{sugaya98}, the effect cannot be explained by the internal motion
of the nucleons in the deuteron.

In contrast,  in \cite{cassing94a}, $\bar{\mrm{p}}$~production in p+A and
d+A reactions is analysed within a phase-space model incorporating the
self-energies  of the baryons.    It  is  claimed  that  the  internal
momentum distribution of  the deuteron provides  a natural explanation
of the large enhancement under discussion.

In  \cite{muellerh04}  we     reproduced the   observed  increase   of
subthreshold $\bar{\mrm{p}}$~production  in dC and $\alpha$C reactions
compared to pC interactions and demonstrated that within the ROC model
the  number  of  participating  nucleons   is  the key  quantity   for
understanding subthreshold particle production. In addition, the whole
set  of  experimental  data  for  all    ejectiles measured  was  well
described.  This is of special importance  since there is not only the
enhancement  of  $\bar{\mrm{p}}$~production, but also  the increase of
the $\mrm{K}^-$~cross sections with  increasing energy and mass number
of  the projectile  around   the  elementary production  threshold  at
2.6~GeV.  Also the  completely  different energy  and projectile  mass
dependence  of  the  pion production  cross   sections  far above  the
threshold could be reproduced.

Here, we repeat these considerations for the  heavier target nuclei Cu
and  Pb    and     come to   rather      similar  consequences.     In
figure~\ref{fCuAll}, the  results of the  ROC  model for pCu, dCu  and
$\alpha$Cu reactions are compared to the data~\cite{sugaya98}.  Again,
the  overall agreement is quite satisfactory  in view of the different
projectile types, the large region of  incidence energies, the variety
of ejectile  species  and the  huge  differences  of   many orders  of
magnitude  in the considered cross section  values.  In case of the Pb
target (see figure \ref{fPbAll})  only data for d-induced reactions up
to 5~A MeV and for  p-induced reactions at  12~GeV are available. Also
for such a heavy target the reproduction of the data is satisfactory.

For 12~GeV protons the   ROC  calculations tend to underestimate   the
K$^\pm$   and  $\bar{\mrm{p}}$  data  (see    figures \ref{fCuAll} and
\ref{fPbAll}).     In    figure~\ref{ITEP}   a    similar    data  set
\cite{vorontsov88} is compared   with  ROC-model  calculations.    The
spectra are measured at a slightly lower energy and at a similar angle
but in a much wider momentum region compared to the KEK data.  In this
case the tendency  of the ROC results goes  in the opposite  direction
and overestimates   the   cross sections  for the  heavier   ejectiles
$\mrm{K}^\pm$ and    $\bar{\mrm{p}}$.  Also for   the  higher ejectile
momenta the description  of the whole  data set for targets between Be
(see figure 3 in  \cite{komarov04}) and Ta is satisfactory. Typically,
the deviations between our  calculations and the  data are less than a
factor of 2.

\begin{table}
   \caption{Calculated      cross   sections  and statistical   errors
   $\sigma_p$, $\sigma_d$ and $\sigma_\alpha$ for $\bar{p}$~production
   in p-, d- and  $\alpha$-induced interactions, respectively, with Cu
   at       3.5  A  GeV   for       different temperature   parameters
   $\Theta^{max}_R$.}

   \label{table} \lineup
   \begin{indented}
   \item[] \begin{tabular}{cccc}
   \br
   $\Theta^{max}_R$(MeV)&$\sigma_p$(nb)&$\sigma_d(\mrm{\mu b})$&
                                     $\sigma_\alpha(\mrm{\mu b})$\\
   \mr
   7.0& 5.2 $\pm$ 0.2 & 0.57 $\pm$ 0.03 & 1.7 $\pm$ 0.1 \\
   7.5& 2.7 $\pm$ 0.1 & 0.39 $\pm$ 0.05 & 1.5 $\pm$ 0.2 \\
   8.0& 1.9 $\pm$ 0.2 & 0.27 $\pm$ 0.02 & 1.2 $\pm$ 0.2 \\
   \br
   \end{tabular}
   \end{indented}
\end{table}

Also an assumed uncertainty  of 0.5~MeV for the  temperature parameter
$\Theta^{max}_R$ yields deviations within this   limit at the   lowest
energy    where  the    enhancement    effect  was   observed     (see
table~\ref{table}).  As  shown in \cite{komarov04}  the dependence  of
the    calculated results  on   $\Theta^{max}_R$  becomes smaller with
increasing projectile energy.   From table~\ref{table} can be  deduced
that at a given incidence energy per  nucleon an increasing projectile
mass  diminishes the dependence   of the subthreshold production cross
section on the temperature parameter  $\Theta^{max}_R$ of the residual
nucleus, too. This can be explained by the  larger effective number of
participants  in  case of  the heavier   projectile, which  leads to an
increase of the energy available for particle production.

A striking  feature of the obtained results  is the consistent quality
of  data reproduction for all types  of final particles independent of
the target mass.  With regard to the fact that  the cross sections for
the  interaction of the  various  final particles with nucleons differ
considerably one  should, especially for   heavy target nuclei, expect
distinct deviations of the calculated spectra  from the measured ones,
since individual properties of the various outgoing particles like the
mentioned cross sections are not (yet) incorporated  in the model.  In
the light of this result the question of  the treatment of final-state
interactions is worth being discussed.   It  is the excitation of  the
residual nucleus which  diminishes the  energy available for  particle
production in the model calculations. This decreases the corresponding
cross sections especially in  the threshold region independent of  the
type of secondary particles.  It seems that individual interactions of
the outgoing particles with the target nucleons do not play a decisive
role. A rigorous explanation would be the assumption  of a finite life
time  of intermediate states called  FBs in the  ROC  model. If such a
life  time would be comparable  to the time the FBs  need to leave the
residual nucleus,   then the  individual   properties  of the  various
particles  would  wash-out.  We    are   aware of the   difficulty  to
articulate such a point of view regarding the fact  that a major topic
of  present  research   activities consists  in  investigating  medium
modifications of  individual particle properties.   On the other hand,
although the  deviations  between our  calculations and the   data are
usually less than a factor of 2 or so, there remains  room for not too
large effects not yet included in the ROC model.

As shown  in \cite{muellerh04} and  supported  by the present results,
the  main contribution   for   subthreshold  particle   production  in
conditions of the experiments  \cite{sugaya98} comes, according to the
ROC model, from the excitation of FBs with baryon numbers from 2 to 4.
This means   that physical  properties of  such  FBs  may be  directly
studied in collisions  of proton, deuteron  and  helium-ion beams with
the lightest  nuclear targets, $^2$H, $^3$He  and $^4$He. In this case
such complicating   effects  like  varying numbers    of participating
nucleons, loss  of energy for the  excitation of the residual nucleus,
final-state  absorption     and rescattering   in   the  residue,  are
significantly  diminished or removed   at  all.  The  model reveals  a
complete   analogy of  the interaction  of  few-nucleon  groups to the
interaction    of single   nucleons   in  the  subthreshold production
processes.   This may be treated  as  an indication that properties of
few-nucleon groups in  conditions of high energy-momentum transfer are
governed by quark-gluon degrees of freedom.  That makes experiments on
subthreshold production at light  target  nuclei a promising tool  for
the study of the transition from hadronic to quark-gluon properties.

\section{Summary \label{sum}}

Subthreshold particle production  is a collective phenomenon  which is
far from being completely  understood.  From the  viewpoint of the ROC
model, data on  subthreshold particle production  can be reproduced by
considering the interaction of  few-nucleon groups in complete analogy
to  the interaction   of single    nucleons,   also with  regard    to
high-momentum transfer processes.   It has been demonstrated here that
in a  wide range of target mass  numbers the  cluster concept yields a
quite natural explanation of  the enhancement of subthreshold particle
production due to the energy  gain in  the interaction of  few-nucleon
groups  compared to NN interactions.  Especially  at energies below or
near  the  NN threshold  the  consideration  of  the interplay between
hadron production and   fragmentation processes is important for   the
reproduction of the cross sections.  This concept should be applicable
not only  in proton-  or   light-ion-induced reactions,  but also  for
heavy-ion interactions,   although in the  latter case  the  number of
partial processes  increases tremendously.   In  this sense,  the  ROC
model can  be  considered  as   a promising   approach to   a  unified
description  of particle production  processes  in a large variety  of
different types of nuclear reactions.

\ack
One of the  authors (H.M.) would like  to thank  W.~Eng\-hardt for the
promotion of this study.

\section*{References}
\bibliographystyle{JourPhysG}
\bibliography{COSY,hadMod,hmbib,hmhelp,nucData,frag,Monte_Carlo}

\end{document}